\documentclass[aps,a4paper,showpacs,showkeys,twocolumn]{revtex4}
\usepackage{amsmath}
\usepackage{amssymb}
\usepackage{graphics}
\usepackage{psfrag}
\usepackage{epsfig}
\usepackage{hyperref}
\usepackage{ulem}
\usepackage{color}

\begin{document}

\title {Role of Frustration in a Weakly Disordered Checkerboard Lattice}
\author{F. M. Zimmer and W. C. Silva}
\affiliation{Instituto de F\'{\i}sica - Universidade Federal de Mato Grosso do Sul, Campo Grande, 79070-900, MS, Brazil}
\author{M. Schmidt}
\affiliation{Departamento de F\'{\i}sica - Universidade Federal de Santa Maria, Santa Maria,  97105-900, RS, Brazil}
\author{S. G. Magalhaes}
\affiliation{Instituto de F\'{\i}sica, Universidade Federal do Rio Grande do Sul, 91501-970,
Porto Alegre, RS, Brazil}

\date{\today}

\begin{abstract}
Quenched disorder effects on frustrated systems are explored by considering random fluctuations  on  the antiferromagnetic (AF) interactions between spins on the checkerboard lattice. 
The replica framework is adopted within a cluster mean-field approach, resulting in an effective single-cluster model. 
This effective model is treated within a one-step replica symmetry breaking (RSB) approach with exact evaluations for all intracluster interactions. 
Competing interactions  are introduced by tuning the ratio $J_2/J_1$ (where $J_1$ and $J_2$ are first-neighbour and  second-neighbor interactions, respectively), which can lead to a highly frustrated scenario when  $J_2/J_1\rightarrow 1$, where a phase transition between AF orders takes place in the absence of disorder. In particular, the AF order appears at lower values of $J_2/J_1$, with the Neel temperature decreasing as the frustration increases. However, quenched disorder changes this description, introducing a RSB spin glass phase for strong enough disorder intensity $J$. 
In fact, for low levels of disorder, a RSB solution with staggered magnetization (mixed phase) emerges from the maximum frustration region.
It suggests that, in the presence of weak quenched disorder, systems with competing interactions are prone to present a glassy behavior instead of conventional orders.

\end{abstract}

\maketitle
\section{Introduction}

Magnetic systems hosting frustration are a continuous source of challenging problems and a platform for novel phenomena. It is well known that the presence of quenched disorder-driven frustration provides a favorable context for the onset of unconventional magnetic phases, such as the spin-glass phase \cite{doi:10.1142/0271, Hertz91, Scripta_Nordblad_SG_Review, Mydosh_Review_2015}. Apart from disordered systems, competing interactions can also lead to a frustrated scenario, which is called simply as frustration from now on. This source of frustration often introduces a competition between magnetic phases \cite{PhysRevE.80.051117, doi:10.1142/5697, FGODOY2020126687}, giving rise to a number of interesting phenomena, such as reentrant transitions and even disestablishing conventional long-range orders \cite{Lacroix_book}. 
From the experimental point of view, frustrated systems seem to be highly sensitive to the presence of disorder. 
This high sensitivity can lead to materials prone to exhibit spin-glass behavior at very low levels of disorder \cite{Gingras2010}. In spite of that, the interplay between disorder and frustration has been explored in a few setups \cite{Mateus2016, Andreanov, yokota, prezimmer2014}. In this context, several efforts have focused on highly frustrated systems, but a relevant issue is whether adjustable levels of frustration can support a spin-glass phase in a scenario of weak quenched disorder. 

Magnetic materials that host both competing interactions and disorder can present rich phase diagrams. For instance, a spin-glass phase between two conventional long-range orders has been reported in the concentration {\it versus} temperature phase diagram of several frustrated magnets. Interesting examples can be found in the pyrochlore antiferromagnet LiGa$_{1-x}$In$_x$Cr$_4$O$_8$ \cite{doi:10.7566/JPSJ.84.043707}, the Kitaev-Heisenberg magnet Ru$_{1-x}$Cr$_x$Cl$_3$ \cite{PhysRevB.99.214410} and the Ising system Fe$_{x}$Mn$_{1-x}$TiO$_{3}$ \cite{doi:10.1143/JPSJ.62.4488}. In particular, the spin-glass phase found near $x=0.5$ in the Fe$_{x}$Mn$_{1-x}$TiO$_{3}$ compound \cite{doi:10.1143/JPSJ.62.4488, doi:10.1143/JPSJ.65.3331, TORIKAI200695} is separated from two different antiferromagnetic orders by mixed-phases, in which spin-glass-like freezing and antiferromagnetic correlations coexist \cite{doi:10.1143/JPSJ.62.4488}. Therefore,  plenty of phenomena can be observed in systems with competing interactions and disorder.

Spin models on bipartite lattices with competing interactions between first-neighbours ($J_1$) and second-neighbours ($J_2$) provide a useful platform to evaluate effects of different degrees of frustration on magnetism. In these systems, one can go from an unfrustrated scenario to a highly frustrated one by tuning the ratio $J_2/J_1$. An interesting example within this class of systems is the Ising model on the checkerboard lattice \cite{PhysRevB.85.134427, PhysRevB.99.144414, sadrzadeh2015phase}. In this lattice, which can be seen as a two-dimensional version of the pyrochlore lattice, the highly frustrated limit is achieved at  $J_2=J_1$, where a ground-state transition between two ordered states takes place.
However, one can already expect frustration effects, such as the reduction in the ordering temperature, when approaching the frustration maximum. 
An interesting question concerns how quenched disorder can affect the coupling-temperature phase diagram of this model. A reasonable posit is that the increase in frustration can lead to a higher sensitivity to perturbations. In this context, one can expect that the conventional long-range orders found near the frustration maximum can be strongly affected by the presence of low levels of disorder.
Motivated by the above issues and by the lack of specific studies of disorder effects on the checkerboard Ising model, we investigate the role of disordered couplings on the Ising checkerboard lattice.

There are a few analytical attempts to deal with quenched disorder and (geometrical) frustration in the same theoretical framework.
For instance, Ref. \cite{Andreanov} considers a strongly frustrated pyrochlore lattice perturbed by weak-exchange randomness to suggest an SG phase transition at low temperature. The findings claim that the freezing temperature is proportional to the disorder strength, without essential deviations from the behavior observed in disordered SG systems without geometrical frustration \cite{Andreanov}. 
In Ref \cite{yokota}, a highly frustrated stacked triangular lattice with randomness in the Ising spin interactions (antiferromagnetic) was studied within a cluster mean-field approach. The results have suggested the presence of SG phase at lower intensities of disorder when compared to the same model with ferromagnetic interactions (without geometrical frustration) \cite{yokota}. 
Another interesting result has been achieved from a cluster formalism that considers several geometrically frustrated clusters with disordered interactions between cluster magnetic moments \cite{prezimmer2014,PhysRevE.89.062117}. In this instance, the findings indicate that the existence of geometrically frustrated clusters potentializes the disordered interaction, leading to a cluster SG phase to appear at lower disorder strength \cite{Schmidt2015416,Mateus2016}. Nevertheless, there is still a lack of results for disordered models in which the degree of frustration can be tuned from an unfrustrated scenario to a highly frustrated regime.

Our approach considers spins on the checkerboard lattice with random fluctuations in the antiferromagnetic interactions: $J_1+\delta J_{ij}$ and  $J_2+\delta J_{ij}$. The fluctuations follow Gaussian probability distributions, introducing a quenched disorder in the problem. We adopt a cluster variational mean-field method in a replica background to take an effective single-cluster model within a one-step replica symmetry breaking (1s-RSB) approach \cite{baviera}. We solve exactly the effective model by considering a random distribution of disorder for the intracluster interactions, and then we perform the average over the intracluster disordered couplings. In this way, the present theoretical framework allows us to evaluate the role of disorder in thermodynamics at different levels of frustration $J_2/J_1$.

The paper is structured as follows. In Section \ref{model} we define the model and the analytical procedure used to get the free energy in the cluster variational mean-field method within the 1s-RSB scheme. In Section \ref{results}, we presented a detailed discussion of the numerical solutions of the free energy and 1s-RSB order parameters in phase diagrams for different configurations of disorder and frustration. The last section \ref{conclusion} is reserved for the conclusions.

\section{Model}\label{model}
We start from the Ising model $H=-\sum_{i,j}J_{ij}\sigma_{i}\sigma_{j}$ with spins  $\sigma_i=\pm 1$ on the site $i$ of the checkerboard lattice with $N$ sites. We adopt AF interactions among first $J_1$ and second-neighbors $J_2$ with random deviations $\delta J_{ij}$. 
The Hamiltonian can be explicitly rewritten as 
\begin{equation}\label{01}
H=-\sum_{\langle i,j\rangle }(J_1+\delta J_{ij})\sigma_{i}\sigma_{j}-\sum_{\langle \langle i,j\rangle \rangle }(J_{2}+\delta J_{ij})\sigma_{i}\sigma_{j},
\end{equation}
where $\langle i,j\rangle$ and $\langle\langle i,j\rangle\rangle$ denote sums over pairs of sites first and second neighbor, respectively. The random deviations follow Gaussian probabilities distributions given by:
\begin{equation}\label{gauss}
P(\delta J_{ij})=\frac{1}{\sqrt{2\pi J^2}}e^{-\frac{{(\delta J_{ij})}^2}{2 J^2}}.
\end{equation}

We can get the thermodynamic behavior of this disordered problem from the configurational average of the free-energy per site: $f= -\frac{1}{\beta N} \overline{\mbox{ln} (Z(\{\delta J_{i,j}\}))}$, in which $\beta=1/T$ ($T$ is the temperature), $Z(\{\delta J_{i,j}\})$ is the partition function for a distribution of $\{\delta J_{i,j}\}$,  and $\overline{\cdots}$ stands for the average over the disorder expressed by Eq. (\ref{gauss}). We handle this interacting problem with a cluster mean-field method (CMF), in which the lattice is divided in $N_c$ clusters with $n_s$ sites each ($N=N_c n_s$). 
In order to clear up the present CMF method, we rewrite the Hamiltonian (\ref{01}) into two parts:
one representing the intracluster interactions $H_{intra}$, and the other describing the inter-cluster interactions $H_{inter}$. It means  $H=H_{intra}+H_{inter}$, where
\begin{equation}\label{intra}
H_{intra}=-\sum_{\nu}^{N_{c}}\sum_{ i_\nu,j_\nu}^{n_s}(J_{i_\nu j_\nu}+\delta J_{i_\nu j_\nu})\sigma_{i_\nu}\sigma_{j_\nu},
\end{equation}
and
\begin{equation}\label{inter}
H_{inter}=-\sum_{\nu,\lambda}^{}\sum_{i_\nu,j_\lambda}(J_{i_\nu,j_\lambda}+\delta J_{i_\nu j_\lambda})\sigma_{i_\nu}\sigma_{j_\lambda},
\end{equation}

with $\nu$ and $\lambda$ denoting cluster labels, and $J_{i_\nu,j_\lambda}=J_1$ or $J_2$. 
In particular, the intercluster interactions are calculated with a mean-field approximation, while the intracluster interactions are evaluated exactly.

\begin{figure}[htbp]
	\centering
	\includegraphics[width=0.7\columnwidth]{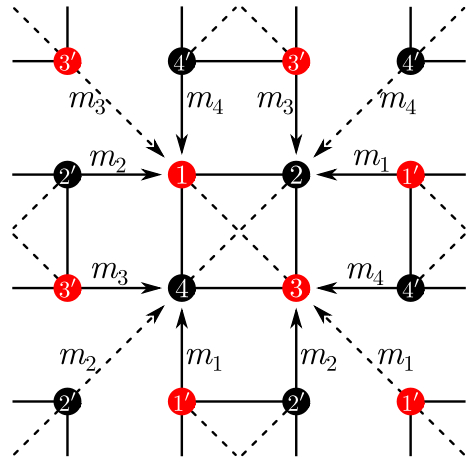}
	\caption{Schematic representation of the checkerboard network divided into clusters with $4$ spins. The white circles (black circles) represent the spins of Ising that assume the value of $-1$ ($1$). The continuous bold lines represent the intracluster interactions and the dashed arrows indicate the effective fields acting on the spins of the border of the central cluster. For the purpose of representation we assume that the system behaves like AF, with the spins $1$ and $3$ ($2$ and $4$) having the same orientation.}\label{fig_clusters}
\end{figure}

We use the replica method to treat the intercluster disorder: $f= -\lim_{N\to \infty} \lim_{n\to 0}\frac{\overline{\ln {\overline{Z^n_{\nu\lambda}}(\{\delta J_{i_\nu,j_\nu }\})}}}{\beta Nn}$, where the intercluster disorder-averaged replicated partition function becomes  
\begin{equation}\begin{split}\label{06}
\overline{Z^n_{\nu\lambda}}(\{\delta J_{i_\nu,j_\nu }\})=
\exp \left[\frac{\beta^{2} J^2 n}{4}\sum_{(i_\nu,j_\lambda)}1\right ]\\
Tr_a \exp \left[-\beta H^{(n)}(\{\delta J_{i_\nu,j_\nu}\})\right]
\end{split}\end{equation} 
with the replicated Hamiltonian
\begin{equation}\begin{split}
H^{(n)}(\{\delta J_{i_\nu,j_\nu}\})=\sum_{a=1}^n 
H_{intra}^a(\{\delta J_{i_\nu,j_\nu}\}) 
\\
-\sum_{(i_\nu,j_\lambda)}[\sum_{a=1}^{n}J_{i_\nu,j_\lambda}\sigma_{i_\nu}^a\sigma_{j_\lambda}^a +{\beta J^2}{}\sum_{a < b}\sigma_{i_\nu}^{a}\sigma_{j_\lambda}^{a}\sigma_{i_\nu}^{b}\sigma_{j_\lambda}^{b}]
\label{h_eff0},
\end{split}\end{equation}
$a$ (or $b$) representing a replica index and $(i_\nu,j_\lambda)$ referring to sums between a site $i_\nu$ in the cluster $\nu$ and its first- or second neighbor site $j_\lambda$ in the cluster $\lambda$.
The free energy can then be recorded as 
\begin{equation}\label{feff}
f= -\frac{3 \beta J^2}{4}-\lim_{n\to 0}\frac{\overline{\ln Tr \exp{[-\beta H^{(n)}(\{\delta J_{i_\nu,j_\nu}\})]}}}{\beta Nn} ,
\end{equation}
which explicitly considers clusters with four sites ($n_s=4$) as depicted in Fig. (\ref{fig_clusters}).

The intercluster interactions are decoupled by adopting a variational approach, 
introducing a cluster mean field treatment \cite{baviera}. 
To be specific, we assume a trial model $\tilde{H}^{(n)}$ that considers a system divided into clusters with the same interactions of Eq. (\ref{h_eff0}) but replacing the intercluster couplings by
\begin{align}\label{parametros2}
\sigma_{i_\nu}^a\sigma_{j_\lambda}^a\sigma_{i_\nu}^b\sigma_{j_\lambda}^b \longrightarrow q^{ab}_{j_\lambda} \sigma_{i_\nu}^a\sigma_{i_\nu}^b+q^{ab}_{i_\nu}\sigma_{j_\lambda}^a\sigma_{j_\lambda}^b\\
\label{parametros3}\sigma_{i_\nu}^{a}\sigma_{j_\lambda}^{a}\longrightarrow \sigma_{i_\nu}^{a}m^a_{j_\lambda}+m^{a}_{i_\nu}\sigma_{j_\lambda}^{a}  ,
\end{align}
where $\{m^{a}_{i_\nu}\}$ and $\{q^{ab}\}$ are sets of variational parameters. Therefore, we get 
\begin{equation}\begin{split}\label{htrial}
\tilde{H}^{(n)}(\{\delta J_{i_\nu,j_\nu}\},m^{a},q^{ab})=\sum_{a=1}^n H_{intra}^a(\{\delta J_{i_\nu,j_\nu}\})+
\\
-\frac{1}{2}\sum_{\nu}\sum_{(i_\nu,j_\lambda)}[\sum_{a}^{n}J_{i_\nu,j_\lambda} m_{j_\lambda}^a\sigma_{i_\nu}^a  +{\beta J^2}{}\sum_{a < b} q^{ab}_{j_\lambda}\sigma_{i_\nu}^{a}\sigma_{i_\nu}^{b}],
\end{split}\end{equation}
where factor 1/2 in the second sum aims to avoid double counting of couplings shared by two different clusters.

This procedure results in the following single-cluster effective problem
\begin{equation}\begin{split}
 f=             \frac{-3\beta J^2 }{4}+\lim_{n\to 0}\frac{1}{n}[\frac{\beta J^2}{4}\sum_{a<b}(q^{ab})^2+ \\+\sum_{a=1}(\sum_{\langle i_\nu,j_\lambda\rangle}\frac{J_1}{2} m^a_{i_\nu}m^a_{j_\lambda} +\sum_{\langle\langle i_\nu,j_\lambda\rangle\rangle}\frac{J_2}{2} m^a_{i_\nu}m^a_{j_\lambda})
\\ 
-\frac{1}{\beta n_s}\overline{\ln {Tr\left\{e^{-\beta H_{eff}^{(n)}} \right\}}}] ,
\end{split}\end{equation}
where $q^{ab}$ and ${m^a}$ provide an extreme for the free energy.

At this time, we use one-step symmetry breaking (1S-RSB) to treat the the variational parameters \cite{parisi}: $m^a=m$ does not depend on the replica index, and $q^{ab}=q_0$ if $I(a/c_1)=I(b/c_1)$ or $q^{ab}=q_1$ if $I(a/c_1)\neq I(b/c_1)$, where $I(x)$ represents the smallest integer greater than or equal to $x$.
Therefore, 
\begin{equation}\begin{split}
f =\frac{3\beta J^2}{4}\left[\left(1-q_{1}\right)^{2}+c_{1}\left(q_{0}^{2}-q_{1}^{2}\right)\right]
\\+\frac{J_{1}}{4}(m_1+m_3)(m_2+m_4)+ 
\frac{J_{2}}{4}(m_{1}m_{3}+m_{2}m_{4}) 
\\ -\frac{1}{4\beta c_{1}}\overline{\int Dz\ln \int Dv \left\{\mbox{Tr } \exp{[-\beta H_{eff}^{RSB}(\{J_{ij}\})]}\right\}^{c_{1}}}, 
\end{split}\label{freeenergy}\end{equation}
 where

 \begin{equation}\begin{split}
 H_{eff}^{RSB}(\{J_{ij}\})  =H_{intra}(\{J_{ij}\})-\sum_{(i,j)}J_{1}\sigma_{i}m_{j}
 \\
-\sum_{((i,j))}J_{2}\sigma_{i}m_{j} -J\sqrt{3}\sum_{i}(\sqrt{q_{0}}z_{i}+\sqrt{q_{1}-q_{0}}v_{i})\sigma_{i},
\end{split}\label{heffective} \end{equation}
 $Dx\equiv\prod_{i=1}^{n_s}dx_i \frac{e^{-x_i^2/2}}{\sqrt{2\pi}}$ ($x=z$ or $v$) and the order parameters $q_0$, $q_1$, $m$ and $c_1$ extremize the free energy, and are explicitly exhibited in Appendix \ref{apendixA}.

 \begin{figure}[htbp]
	\centering
	\includegraphics[width=1\columnwidth]{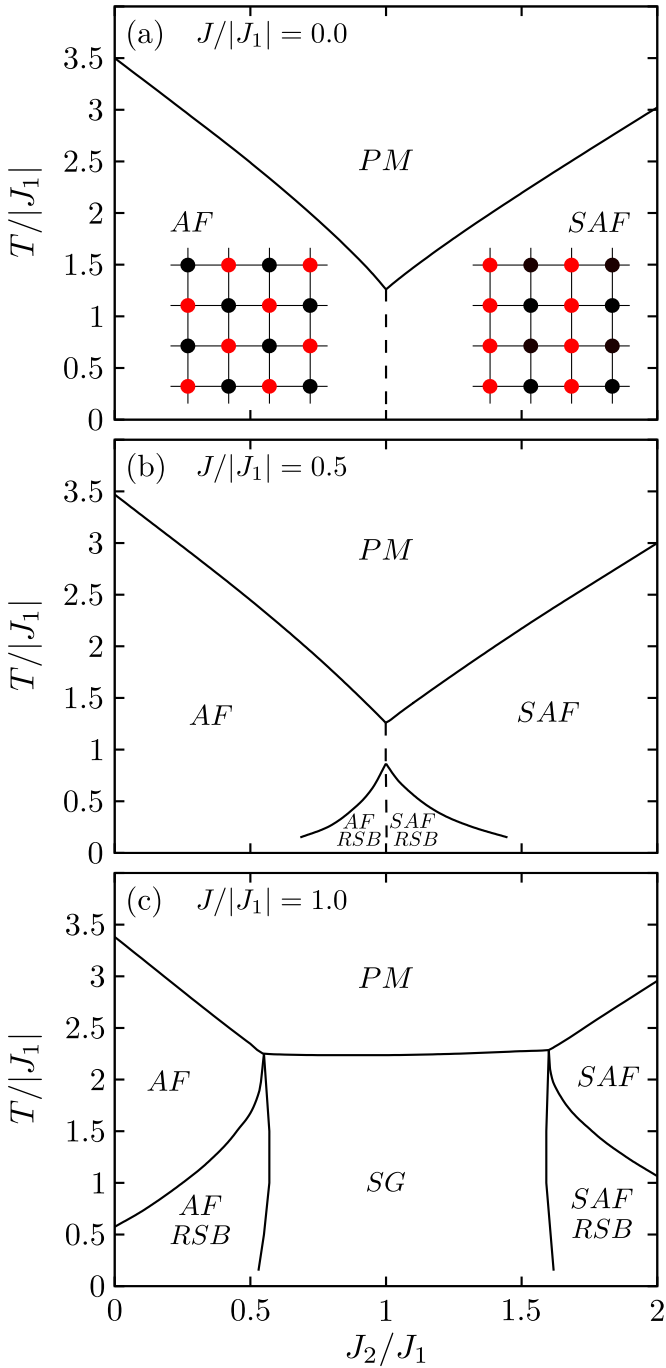}
	\caption{Phase diagrams $T/|J_1|$ versus $J_2/J_1$ for different disorder intensities $J/|J_1|$: 0.0, 0.5, and 1.0 (panels (a), (b), and (c), respectively).}\label{fig_t_vs_r}
\end{figure}

\section{Results}\label{results}

In the following, we present our findings in phase diagrams that explore different scenarios for the role of disorder ($J \geq 0$) on the checkerboard lattice with AF interactions.
We solve self-consistently the set of Eqs. (\ref{a1}), (\ref{a2}), (\ref{a3}), and (\ref{a4}) in order to obtain the free energy, given by Eq. (\ref{freeenergy}).
The SG phase occurs for RSB order parameter $\delta=q_1-q_0 >0$ with zero local magnetizations. The AF and SAF orders are characterized by RS solution ($\delta=0$) with  $m_{AF}=|m_1-m_2+m_3-m_4|/4>0$ (following Fig. \ref{fig_clusters} site numbers) and $m_{SAF}=(|m_1-m_3|+|m_2-m_4|)/4>0$, respectively. We also find a mixed phase, in which the RSB occurs with a finite staggered magnetization $m_{AF}$ (AF RSB) or $m_{SAF}$
(SAF RSB). 

Figure \ref{fig_t_vs_r}(a) shows the temperature {\it versus} frustration parameter $J_2/J_1$ phase diagram for $J=0$. 
In this clean limit, an increase in $J_2$ enhances the AF ground-state energy per spin, which is given by $u_{AF}=-2J_1 + J_2$.
For $J_2/J_1=1$, the system can be found in any state in which the sum of magnetic moments within the squares with crossing interactions is zero \cite{moessner2004planar, PhysRevB.63.224401}. 
For $J_2/J_1>1$, the degeneracy is reduced, but the system can be found in any state composed of antiferromagnetic $J_2$ diagonal chains. In this case, different ground states can be achieved by flipping magnetic moments within $J_2$ diagonals \cite{PhysRevB.85.134427}. Therefore, the second neighbour couplings are fully satisfied while only half the first-neighbour couplings are satisfied.
Within our CMF calculations, the ground-state degeneracy is broken and we consider one of the many possible ground states, which is shown in  the right-hand inset of Fig. \ref{fig_t_vs_r}(a). This phase, hereafter called superantiferromagnetic (SAF), has a ground-state energy $ u_{SAF}= - J_2$ and is present even when thermal fluctuations take place. Therefore, in our CMF phase diagram, the system exhibits a zero-temperature phase transition at $J_2/J_1=1$, as expected. In addition, first order phase transitions between AF and SAF phases take place at finite temperatures. Moreover, the thermal fluctuations can drive second-order phase transitions between the low temperature phases and the PM state. It is worth to note that the critical temperatures of these order-disorder transitions are reduced when $J_2/J_1 \to 1$, which indicates the onset of stronger frustration effects as this limit case is approached from $J_2/J_1=0$ or $J_2/J_1\to \infty$. These findings indicate that $J_2/J_1=1$ corresponds to the maximum of frustration introduced by the competing couplings $J_1$ and $J_2$. Therefore, our CMF calculations are able to incorporate important frustration effects of the model. 

\begin{figure}[t]
	\centering
	\includegraphics[width=0.9\columnwidth]{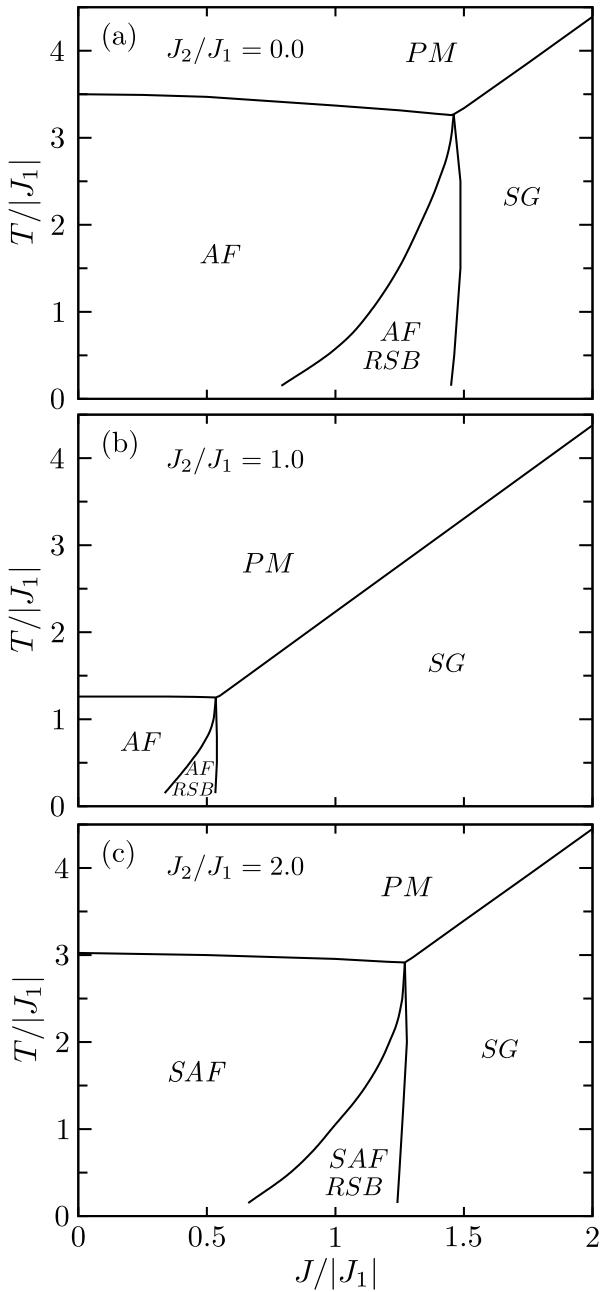}
	\caption{Phase diagrams of temperature as a function of quenched disorder for different levels of frustration: (a) $J_2/J_1=0.0$, (b) 1.0, and (c) 2.0.}\label{figure3}
\end{figure}

Quenched disorder brings an additional source of frustration in the checkerboard lattice. In addition to the competing couplings ($J_1$ and $J_2$), a finite $J$ introduces randomness in the interactions,  causing random deviations in the AF couplings, which can drive relevant changes in the coupling-temperature phase diagram of the model. In Fig. \ref{fig_t_vs_r}(b), we present the phase diagram for $J/|J_1| = 0.5$. In this weakly disordered case, the onset of mixed phases, in which long-range orders coexist with RSB, is found at low temperatures. More importantly, the mixed phase is only found near $J_2/J_1=1$, being absent for $J_2/J_1=0$ and $J_2/J_1=2$. Furthermore, the transition temperature to the RSB solution increases toward $J_2/J_1\rightarrow 1$. Therefore, our findings support that a weak quenched disorder favors a RSB phase in the scenario introduced by the competitive couplings $J_1$ and $J_2$ near the frustration maximum.
In addition, when the quenched disorder enhances, the SG phase is found and becomes dominant around $J_2/J_1=1$, as depicted in Fig. \ref{fig_t_vs_r}(c).   The freezing temperature turns almost independent of $J_2/J_1$, with the AF and SAF  orders only appearing far from $J_2/J_1=1$, being separated by the mixed and SG phases.
It reinforces that the RSB solution grows from the strong-competitive region as the disorder increases. 

Remarkably, the structure of the phase diagram shown in Fig. \ref{fig_t_vs_r}(c) resembles the temperature versus $x$ phase diagrams obtained for  Fe$_{x}$Mn$_{1-x}$TiO$_{3}$ (see Fig. 10 in Ref. \cite{doi:10.1143/JPSJ.62.4488}). In this Ising system, antiferromagnetic long-range orders take place for $x \approx 0$ and $x \approx 1$. However, within the hexagonal planes formed by iron and manganese ions, the interactions between Fe ions are ferromagnetic while the interactions between Mn ions are antiferromagnetic. Therefore, at intermediary concentrations, a competitive scenario driven by the exchange interactions takes place. As a consequence, a spin-glass state is found at $x=0.5$ \cite{doi:10.1143/JPSJ.62.4488, doi:10.1143/JPSJ.66.3636,TORIKAI200695, PhysRevB.43.8199}. In addition, this SG state is separated from the AF orders by two reentrant mixed phases, in which signatures of both spin-glass freezing and AF long-range orders can be spotted. Therefore, not only the structure of the $T-x$ phase diagram, but also the content of the magnetic phases found in Fe$_{x}$Mn$_{1-x}$TiO$_{3}$, resemble the ones found for the $T-J_2/J_1$ phase diagram of the disordered checkerboard lattice. It is worth noting that the competing interactions and the lattice in the present model are notably different of the competing magnetic couplings and crystalline structure found in Fe$_{x}$Mn$_{1-x}$TiO$_{3}$. However, the coupling ratio ($J_2/J_1$) allows us to tune the degree of competition between interactions and, therefore, plays a similar role as the iron concentration in Fe$_{x}$Mn$_{1-x}$TiO$_{3}$. Therefore, our model is able to incorporate relevant ingredients in spin glasses with competing interactions.

In Fig. \ref{figure3}, we present the temperature ($T/|J_1|$) { \it versus} disorder ($J/|J_1|$) phase diagrams at different levels of frustration ($J_2/J_1$). At low levels of disorder, the increase in $J$ reduces the ordering temperature of the conventional long-range orders (AF and SAF). It is worth to note that this finding is in accordance with the experimental results for several disordered magnetic materials. Therefore, the theoretical framework provides an improvement over the canonical mean-field calculations for the Edwards-Anderson model.
Figure \ref{figure3}(c) exhibits an analogous case for $J_2/J_1=2$, where the SAF order appears at low disorder levels with the SG dominating the phase diagram only at strong disorder intensities ($J/|J_1|>1.3$).
To summarize, if frustration is entirely from disorder, the nontrivial RSB solution comes out only at higher disorder levels.
On the other hand, Fig.  \ref{figure3}(b) shows the SG phase occurring at lower levels of disorder at $J_2/J_1=1$. In this case, the AF order is the stable one for infinitesimal disorders, but the RSB solution occurs for small values of $J$. By comparing the phase diagrams of Fig. \ref{figure3}, it becomes clear that the SG phase can appear at lower levels of disorder when frustration is maximum.  It means that the SG phase is favored against the AF (or SAF) order at lower disorder intensities when the AF competing interactions also introduce frustration.

\begin{figure}[t]
	\centering
	\includegraphics[width=1.\columnwidth]{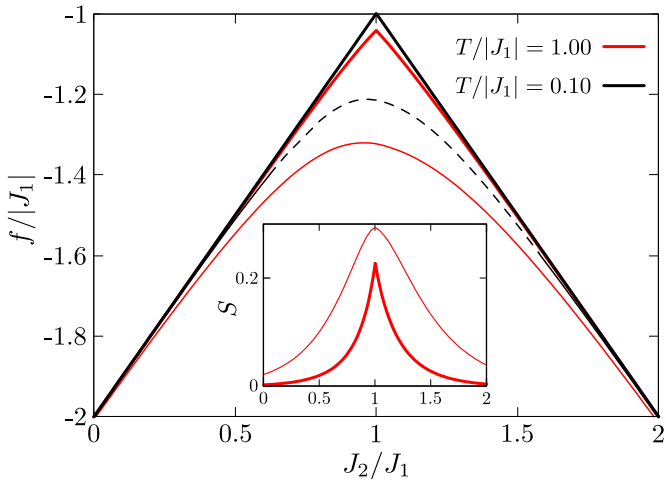}
	\caption{Free energy as a function of frustration parameter for constant temperatures, $T/|J_1|=1.00$ and $0.10$, and different disorder intensities: $J/|J_1|=0.0$ (thick line) and 0.5 (thin line) . The dashed line denotes 1s-RSB solution when $T/|J_1|=0.10$ with  $J/|J_1|=0.50$. The inset exhibits the entropy behavior for $T/|J_1|=1.00$ and the two intensities of disorder ($J/|J_1|=0.0$ and 0.5). }\label{figure4}
\end{figure}

In order to discuss the occurrence of the RSB solution at lower levels of disorder, we present the free-energy as a function of the frustration in Fig. \ref{figure4}.
The thick lines present results for the clean limit ($J=0$), in which the free-energy of the ordered phase (AF or SAF) becomes higher when frustration is increased ($J_2/J_1 \to 1$). It means that the frustration leads to a high free-energy ordered phase. In addition, there is a sharp discontinuity at the AF/SAF phase transition.
When quenched disorder is introduced, the system free-energy decreases significantly near the frustration maximum (see thin lines of Fig. \ref{figure4} for $J/|J_1|=0.50$). 
Therefore, the instability taking place for $J_2/J_1 \to 1$ creates conditions that favor the emergence of states related to the quenched disorder such as the mixed phase.
Furthermore, the discontinuity at the transition between the mixed phases  AF-RSB and SAF-RSB is smooth as compared to the AF/SAF transition (without disorder). The inset brings the entropy behavior for a constant temperature ($T/|J_1|=1.0$) in the AF and SAF phases. The entropy reaches its maximum value at the maximum frustration range, with a further increasing in the presence of quenched disorder (thin line). 
This entropy increasing can also destabilize the "pure" AF (or SAF) order (RS stable solution), helping the mixed phase (RSB solution) to be found at lower disordered strengths in the presence of frustration. It suggests that, at low enough temperatures, RSB takes place within the highly entropic ordered phase, driving an entropy release.

\section{Conclusion}\label{conclusion}

We study the effect of quenched disorder in the antiferromagnetic Ising spin model on the checkerboard lattice.
The problem is dealt with a replica-cluster mean-field formalism, leading to an effective cluster model which is solved exactly with 1s-RSB.
In this model, first ($J_1$) and second-neighbor ($J_2$) AF interactions can lead to a competing situation that can also be affected by random deviations coming from disorder $J$. In this way, we can assess the interplay between frustration coming from different sources (the AF competing interactions and the quenched disorder) on the glassy behavior.

In the clean disorder limit $J=0$, we obtain AF and SAF ground state orders with a discontinuous transition between them at $J_2/J_1=1$, where frustration is maximum. The thermal fluctuations lead to continuous transitions to a PM phase at the Neel temperature, which reaches its minimum value at $J_2/J_1\approx1$. It means that frustration is against conventional orders, destabilizing the antiferromagnetism and helping the PM phase to appear at lower temperatures. In other words, the free energy of ordered phases increases as $J_2/J_1\to1$. 
For strong disordered regimes, the SG RSB solution replaces the antiferromagnetic phases, dominating the phase diagram. However, an interesting phenomenon occurs in small disorder intensities when the competitive scenario is enhanced. In this frustrated regime, the RSB solution appears as a mixed phase. It means that in this competitive scenario RSB takes place even at low levels of disorder. 

To conclude, magnetic systems with competing interactions can present a stronger sensitivity to disorder, which can favor a spin-glass phase in between two competing long-range orders. We also suggest that even when frustration is not strong enough to avoid an ordered phase, it can still favor the onset of a glassy phase at low levels of disorder. This phase can occur as a reentrant mixed spin-glass phase or a canonical spin-glass,  depending on the subtle balance of disorder and frustration of the particular system.

\appendix
\section{Order Parameters}\label{apendixA}
By extremizing the free energy given by Eq. (\ref{freeenergy}), we obtain the following set of equations for the order parameters: 
\begin{equation}
m_{k}= \overline{ \int Dz \left(
\frac{\int Dv[Z_{eff}(J_{ij}) ]^{c_1-1} \langle\sigma_k\rangle_{Z_{eff}}}{\int Dv [Z_{eff}(J_{ij})]^{c_1}} \right)}^{J_{ij}}
\label{a1},\end{equation}
\begin{equation}
q_{0}=\frac{1}{4} \sum_{k=1}^{4} \overline{ \int Dz 
\left(\frac{\int Dv[Z_{eff}(J_{ij}) ]^{c_1-1} \langle\sigma_k\rangle_{Z_{eff}}}{\int Dv [Z_{eff}(J_{ij})]^{c_1}}\right)^2 }^{J_{ij}}
\label{a2},\end{equation}

\begin{equation}
q_{1}=\frac{1}{4} \sum_{k=1}^{4} \overline{ \int Dz 
\frac{\int Dv[Z_{eff}(J_{ij}) ]^{c_1-2} \langle\sigma_k\rangle_{Z_{eff}}^2}{\int Dv [Z_{eff}(J_{ij})]^{c_1}} }^{J_{ij}}
\label{a3},\end{equation}
and
\begin{equation}\begin{split}
3 (J\beta)^2(q_1^2-q_0^2)+\frac{1}{c_1^2}\overline{\int Dz \ln\int Dv[Z_{eff}(J_{ij})]^{c_1}}^{J_{ij}}
\\
-\frac{1}{c_1}\overline{\frac{\int Dv[Z_{eff}(J_{ij}) ]^{c_1} \ln{Z_{eff}(J_{ij})}}{\int Dv [Z_{eff}(J_{ij})]^{c_1}} }^{J_{ij}}=0\end{split}\label{a4}\end{equation}
with $k= 1 \cdots 4$, $Z_{eff}(J_{ij})=\mbox{Tr} \exp{(-\beta H_{eff}(J_{ij}))}$ and $ \langle\sigma_k\rangle_{Z_{eff}}=\mbox{Tr} ~\sigma_k\exp{(-\beta H_{eff}(J_{ij}))}$, with $H_{eff}(J_{ij})$ defined in Eq. (\ref{heffective}).
 
\subsection*{Acknowledgments}
FMZ and SGM acknowledge the support from CNPq/Brazil and WCS thanks gratefully to CAPES/Brazil. MS acknowledges the support of Funda\c{c}\~ao de Amparo \`a Pesquisa do Estado do Rio Grande do Sul (Fapergs).

\bibliography{References}{}

\begin{thebibliography}{29}
\expandafter\ifx\csname natexlab\endcsname\relax\def\natexlab#1{#1}\fi
\expandafter\ifx\csname bibnamefont\endcsname\relax
  \def\bibnamefont#1{#1}\fi
\expandafter\ifx\csname bibfnamefont\endcsname\relax
  \def\bibfnamefont#1{#1}\fi
\expandafter\ifx\csname citenamefont\endcsname\relax
  \def\citenamefont#1{#1}\fi
\expandafter\ifx\csname url\endcsname\relax
  \def\url#1{\texttt{#1}}\fi
\expandafter\ifx\csname urlprefix\endcsname\relax\def\urlprefix{URL }\fi
\providecommand{\bibinfo}[2]{#2}
\providecommand{\eprint}[2][]{\url{#2}}

\bibitem[{\citenamefont{Mezard et~al.}(1986)\citenamefont{Mezard, Parisi, and
  Virasoro}}]{doi:10.1142/0271}
\bibinfo{author}{\bibfnamefont{M.}~\bibnamefont{Mezard}},
  \bibinfo{author}{\bibfnamefont{G.}~\bibnamefont{Parisi}}, \bibnamefont{and}
  \bibinfo{author}{\bibfnamefont{M.}~\bibnamefont{Virasoro}},
  \emph{\bibinfo{title}{Spin Glass Theory and Beyond}}
  (\bibinfo{publisher}{WORLD SCIENTIFIC}, \bibinfo{year}{1986}).

\bibitem[{\citenamefont{Fischer and Hertz.}(1993)}]{Hertz91}
\bibinfo{author}{\bibfnamefont{K.~H.} \bibnamefont{Fischer}} \bibnamefont{and}
  \bibinfo{author}{\bibfnamefont{J.~A.} \bibnamefont{Hertz.}},
  \emph{\bibinfo{title}{{Spin Glasses}}} (\bibinfo{publisher}{Cambridge
  University Press}, \bibinfo{year}{1993}).

\bibitem[{\citenamefont{Nordblad}(2013)}]{Scripta_Nordblad_SG_Review}
\bibinfo{author}{\bibfnamefont{P.}~\bibnamefont{Nordblad}},
  \bibinfo{journal}{Physica Scripta} \textbf{\bibinfo{volume}{88}},
  \bibinfo{pages}{058301} (\bibinfo{year}{2013}).

\bibitem[{\citenamefont{Mydosh}(2015)}]{Mydosh_Review_2015}
\bibinfo{author}{\bibfnamefont{J.~A.} \bibnamefont{Mydosh}},
  \bibinfo{journal}{Reports on Progress in Physics}
  \textbf{\bibinfo{volume}{78}}, \bibinfo{pages}{052501}
  (\bibinfo{year}{2015}).

\bibitem[{\citenamefont{Yin and Landau}(2009)}]{PhysRevE.80.051117}
\bibinfo{author}{\bibfnamefont{J.}~\bibnamefont{Yin}} \bibnamefont{and}
  \bibinfo{author}{\bibfnamefont{D.~P.} \bibnamefont{Landau}},
  \bibinfo{journal}{Phys. Rev. E} \textbf{\bibinfo{volume}{80}},
  \bibinfo{pages}{051117} (\bibinfo{year}{2009}).

\bibitem[{\citenamefont{Diep}(2005)}]{doi:10.1142/5697}
\bibinfo{author}{\bibfnamefont{H.~T.} \bibnamefont{Diep}},
  \emph{\bibinfo{title}{Frustrated Spin Systems}} (\bibinfo{publisher}{WORLD
  SCIENTIFIC}, \bibinfo{year}{2005}).

\bibitem[{\citenamefont{Godoy et~al.}(2020)\citenamefont{Godoy, Schmidt, and
  Zimmer}}]{FGODOY2020126687}
\bibinfo{author}{\bibfnamefont{P.~F.} \bibnamefont{Godoy}},
  \bibinfo{author}{\bibfnamefont{M.}~\bibnamefont{Schmidt}}, \bibnamefont{and}
  \bibinfo{author}{\bibfnamefont{F.~M.} \bibnamefont{Zimmer}},
  \bibinfo{journal}{Physics Letters A} \textbf{\bibinfo{volume}{384}},
  \bibinfo{pages}{126687} (\bibinfo{year}{2020}), ISSN
  \bibinfo{issn}{0375-9601}.

\bibitem[{\citenamefont{Lacroix et~al.}(2011)\citenamefont{Lacroix, Mendels,
  and Mila}}]{Lacroix_book}
\bibinfo{editor}{\bibfnamefont{C.}~\bibnamefont{Lacroix}},
  \bibinfo{editor}{\bibfnamefont{P.}~\bibnamefont{Mendels}}, \bibnamefont{and}
  \bibinfo{editor}{\bibfnamefont{F.}~\bibnamefont{Mila}}, eds.,
  \emph{\bibinfo{title}{{Introduction to Frustrated Magnetism}}}
  (\bibinfo{publisher}{Springer}, \bibinfo{year}{2011}), ISBN
  \bibinfo{isbn}{3642105882}.

\bibitem[{\citenamefont{Gardner et~al.}(2010)\citenamefont{Gardner, Gingras,
  and Greedan}}]{Gingras2010}
\bibinfo{author}{\bibfnamefont{J.~S.} \bibnamefont{Gardner}},
  \bibinfo{author}{\bibfnamefont{M.~J.~P.} \bibnamefont{Gingras}},
  \bibnamefont{and} \bibinfo{author}{\bibfnamefont{J.~E.}
  \bibnamefont{Greedan}}, \bibinfo{journal}{Rev. Mod. Phys.}
  \textbf{\bibinfo{volume}{82}}, \bibinfo{pages}{53} (\bibinfo{year}{2010}).

\bibitem[{\citenamefont{Schmidt et~al.}(2017)\citenamefont{Schmidt, Zimmer, and
  Magalhaes}}]{Mateus2016}
\bibinfo{author}{\bibfnamefont{M.}~\bibnamefont{Schmidt}},
  \bibinfo{author}{\bibfnamefont{F.~M.} \bibnamefont{Zimmer}},
  \bibnamefont{and} \bibinfo{author}{\bibfnamefont{S.~G.}
  \bibnamefont{Magalhaes}}, \bibinfo{journal}{J. Phys.: Condens. Matter}
  \textbf{\bibinfo{volume}{29}}, \bibinfo{pages}{165801}
  (\bibinfo{year}{2017}).

\bibitem[{\citenamefont{Andreanov et~al.}(2010)\citenamefont{Andreanov,
  Chalker, Saunders, and Sherrington}}]{Andreanov}
\bibinfo{author}{\bibfnamefont{A.}~\bibnamefont{Andreanov}},
  \bibinfo{author}{\bibfnamefont{J.~T.} \bibnamefont{Chalker}},
  \bibinfo{author}{\bibfnamefont{T.~E.} \bibnamefont{Saunders}},
  \bibnamefont{and}
  \bibinfo{author}{\bibfnamefont{D.}~\bibnamefont{Sherrington}},
  \bibinfo{journal}{Phys. Rev. B} \textbf{\bibinfo{volume}{81}},
  \bibinfo{pages}{014406} (\bibinfo{year}{2010}),
  \urlprefix\url{https://link.aps.org/doi/10.1103/PhysRevB.81.014406}.

\bibitem[{\citenamefont{Yokota}(2014)}]{yokota}
\bibinfo{author}{\bibfnamefont{T.}~\bibnamefont{Yokota}},
  \bibinfo{journal}{Phys. Rev. E} \textbf{\bibinfo{volume}{89}},
  \bibinfo{pages}{012128} (\bibinfo{year}{2014}),
  \urlprefix\url{https://link.aps.org/doi/10.1103/PhysRevE.89.012128}.

\bibitem[{\citenamefont{Zimmer et~al.}(2014{\natexlab{a}})\citenamefont{Zimmer,
  Silva, Magalhaes, and Lacroix}}]{prezimmer2014}
\bibinfo{author}{\bibfnamefont{F.~M.} \bibnamefont{Zimmer}},
  \bibinfo{author}{\bibfnamefont{C.~F.} \bibnamefont{Silva}},
  \bibinfo{author}{\bibfnamefont{S.~G.} \bibnamefont{Magalhaes}},
  \bibnamefont{and} \bibinfo{author}{\bibfnamefont{C.}~\bibnamefont{Lacroix}},
  \bibinfo{journal}{Phys. Rev. E} \textbf{\bibinfo{volume}{89}},
  \bibinfo{pages}{022120} (\bibinfo{year}{2014}{\natexlab{a}}).

\bibitem[{\citenamefont{Okamoto et~al.}(2015)\citenamefont{Okamoto, Nilsen,
  Nakazono, and Hiroi}}]{doi:10.7566/JPSJ.84.043707}
\bibinfo{author}{\bibfnamefont{Y.}~\bibnamefont{Okamoto}},
  \bibinfo{author}{\bibfnamefont{G.~J.} \bibnamefont{Nilsen}},
  \bibinfo{author}{\bibfnamefont{T.}~\bibnamefont{Nakazono}}, \bibnamefont{and}
  \bibinfo{author}{\bibfnamefont{Z.}~\bibnamefont{Hiroi}},
  \bibinfo{journal}{Journal of the Physical Society of Japan}
  \textbf{\bibinfo{volume}{84}}, \bibinfo{pages}{043707}
  (\bibinfo{year}{2015}).

\bibitem[{\citenamefont{Bastien et~al.}(2019)\citenamefont{Bastien, Roslova,
  Haghighi, Mehlawat, Hunger, Isaeva, Doert, Vojta, B\"uchner, and
  Wolter}}]{PhysRevB.99.214410}
\bibinfo{author}{\bibfnamefont{G.}~\bibnamefont{Bastien}},
  \bibinfo{author}{\bibfnamefont{M.}~\bibnamefont{Roslova}},
  \bibinfo{author}{\bibfnamefont{M.~H.} \bibnamefont{Haghighi}},
  \bibinfo{author}{\bibfnamefont{K.}~\bibnamefont{Mehlawat}},
  \bibinfo{author}{\bibfnamefont{J.}~\bibnamefont{Hunger}},
  \bibinfo{author}{\bibfnamefont{A.}~\bibnamefont{Isaeva}},
  \bibinfo{author}{\bibfnamefont{T.}~\bibnamefont{Doert}},
  \bibinfo{author}{\bibfnamefont{M.}~\bibnamefont{Vojta}},
  \bibinfo{author}{\bibfnamefont{B.}~\bibnamefont{B\"uchner}},
  \bibnamefont{and} \bibinfo{author}{\bibfnamefont{A.~U.~B.}
  \bibnamefont{Wolter}}, \bibinfo{journal}{Phys. Rev. B}
  \textbf{\bibinfo{volume}{99}}, \bibinfo{pages}{214410}
  (\bibinfo{year}{2019}),
  \urlprefix\url{https://link.aps.org/doi/10.1103/PhysRevB.99.214410}.

\bibitem[{\citenamefont{Katori and Ito}(1993)}]{doi:10.1143/JPSJ.62.4488}
\bibinfo{author}{\bibfnamefont{H.~A.} \bibnamefont{Katori}} \bibnamefont{and}
  \bibinfo{author}{\bibfnamefont{A.}~\bibnamefont{Ito}},
  \bibinfo{journal}{Journal of the Physical Society of Japan}
  \textbf{\bibinfo{volume}{62}}, \bibinfo{pages}{4488} (\bibinfo{year}{1993}).

\bibitem[{\citenamefont{Ito et~al.}(1996)\citenamefont{Ito, Morimoto, and
  Aruga~Katori}}]{doi:10.1143/JPSJ.65.3331}
\bibinfo{author}{\bibfnamefont{A.}~\bibnamefont{Ito}},
  \bibinfo{author}{\bibfnamefont{S.}~\bibnamefont{Morimoto}}, \bibnamefont{and}
  \bibinfo{author}{\bibfnamefont{H.}~\bibnamefont{Aruga~Katori}},
  \bibinfo{journal}{Journal of the Physical Society of Japan}
  \textbf{\bibinfo{volume}{65}}, \bibinfo{pages}{3331} (\bibinfo{year}{1996}).

\bibitem[{\citenamefont{Torikai et~al.}(2006)\citenamefont{Torikai, Ito,
  Watanabe, and Nagamine}}]{TORIKAI200695}
\bibinfo{author}{\bibfnamefont{E.}~\bibnamefont{Torikai}},
  \bibinfo{author}{\bibfnamefont{A.}~\bibnamefont{Ito}},
  \bibinfo{author}{\bibfnamefont{I.}~\bibnamefont{Watanabe}}, \bibnamefont{and}
  \bibinfo{author}{\bibfnamefont{K.}~\bibnamefont{Nagamine}},
  \bibinfo{journal}{Physica B: Condensed Matter}
  \textbf{\bibinfo{volume}{374-375}}, \bibinfo{pages}{95}
  (\bibinfo{year}{2006}), ISSN \bibinfo{issn}{0921-4526},
  \bibinfo{note}{proceedings of the Tenth International Conference on Muon Spin
  Rotation, Relaxation and Resonance},
  \urlprefix\url{https://www.sciencedirect.com/science/article/pii/S0921452605012238}.

\bibitem[{\citenamefont{Henry et~al.}(2012)\citenamefont{Henry, Holdsworth,
  Mila, and Roscilde}}]{PhysRevB.85.134427}
\bibinfo{author}{\bibfnamefont{L.-P.} \bibnamefont{Henry}},
  \bibinfo{author}{\bibfnamefont{P.~C.~W.} \bibnamefont{Holdsworth}},
  \bibinfo{author}{\bibfnamefont{F.}~\bibnamefont{Mila}}, \bibnamefont{and}
  \bibinfo{author}{\bibfnamefont{T.}~\bibnamefont{Roscilde}},
  \bibinfo{journal}{Phys. Rev. B} \textbf{\bibinfo{volume}{85}},
  \bibinfo{pages}{134427} (\bibinfo{year}{2012}),
  \urlprefix\url{https://link.aps.org/doi/10.1103/PhysRevB.85.134427}.

\bibitem[{\citenamefont{Sadrzadeh et~al.}(2019)\citenamefont{Sadrzadeh,
  Haghshenas, and Langari}}]{PhysRevB.99.144414}
\bibinfo{author}{\bibfnamefont{M.}~\bibnamefont{Sadrzadeh}},
  \bibinfo{author}{\bibfnamefont{R.}~\bibnamefont{Haghshenas}},
  \bibnamefont{and} \bibinfo{author}{\bibfnamefont{A.}~\bibnamefont{Langari}},
  \bibinfo{journal}{Phys. Rev. B} \textbf{\bibinfo{volume}{99}},
  \bibinfo{pages}{144414} (\bibinfo{year}{2019}),
  \urlprefix\url{https://link.aps.org/doi/10.1103/PhysRevB.99.144414}.

\bibitem[{\citenamefont{Sadrzadeh and Langari}(2015)}]{sadrzadeh2015phase}
\bibinfo{author}{\bibfnamefont{M.}~\bibnamefont{Sadrzadeh}} \bibnamefont{and}
  \bibinfo{author}{\bibfnamefont{A.}~\bibnamefont{Langari}},
  \bibinfo{journal}{The European Physical Journal B}
  \textbf{\bibinfo{volume}{88}}, \bibinfo{pages}{1} (\bibinfo{year}{2015}).

\bibitem[{\citenamefont{Zimmer et~al.}(2014{\natexlab{b}})\citenamefont{Zimmer,
  Schmidt, and Magalhaes}}]{PhysRevE.89.062117}
\bibinfo{author}{\bibfnamefont{F.~M.} \bibnamefont{Zimmer}},
  \bibinfo{author}{\bibfnamefont{M.}~\bibnamefont{Schmidt}}, \bibnamefont{and}
  \bibinfo{author}{\bibfnamefont{S.~G.} \bibnamefont{Magalhaes}},
  \bibinfo{journal}{Phys. Rev. E} \textbf{\bibinfo{volume}{89}},
  \bibinfo{pages}{062117} (\bibinfo{year}{2014}{\natexlab{b}}).

\bibitem[{\citenamefont{Schmidt et~al.}(2015)\citenamefont{Schmidt, Zimmer, and
  Magalhaes}}]{Schmidt2015416}
\bibinfo{author}{\bibfnamefont{M.}~\bibnamefont{Schmidt}},
  \bibinfo{author}{\bibfnamefont{F.}~\bibnamefont{Zimmer}}, \bibnamefont{and}
  \bibinfo{author}{\bibfnamefont{S.}~\bibnamefont{Magalhaes}},
  \bibinfo{journal}{Physica A: Statistical Mechanics and its Applications}
  \textbf{\bibinfo{volume}{438}}, \bibinfo{pages}{416} (\bibinfo{year}{2015}),
  ISSN \bibinfo{issn}{0378-4371}.

\bibitem[{\citenamefont{Baviera et~al.}(1998)\citenamefont{Baviera, Pasquini,
  and Serva}}]{baviera}
\bibinfo{author}{\bibfnamefont{R.}~\bibnamefont{Baviera}},
  \bibinfo{author}{\bibfnamefont{M.}~\bibnamefont{Pasquini}}, \bibnamefont{and}
  \bibinfo{author}{\bibfnamefont{M.}~\bibnamefont{Serva}},
  \textbf{\bibinfo{volume}{31}}, \bibinfo{pages}{4127} (\bibinfo{year}{1998}),
  \urlprefix\url{https://doi.org/10.1088/0305-4470/31/18/005}.

\bibitem[{\citenamefont{Parisi}(1980)}]{parisi}
\bibinfo{author}{\bibfnamefont{G.}~\bibnamefont{Parisi}},
  \bibinfo{journal}{Journal of Physics A: Mathematical and General}
  \textbf{\bibinfo{volume}{13}}, \bibinfo{pages}{1101} (\bibinfo{year}{1980}).

\bibitem[{\citenamefont{Moessner et~al.}(2004)\citenamefont{Moessner,
  Tchernyshyov, and Sondhi}}]{moessner2004planar}
\bibinfo{author}{\bibfnamefont{R.}~\bibnamefont{Moessner}},
  \bibinfo{author}{\bibfnamefont{O.}~\bibnamefont{Tchernyshyov}},
  \bibnamefont{and} \bibinfo{author}{\bibfnamefont{S.~L.}
  \bibnamefont{Sondhi}}, \bibinfo{journal}{Journal of statistical physics}
  \textbf{\bibinfo{volume}{116}}, \bibinfo{pages}{755} (\bibinfo{year}{2004}).

\bibitem[{\citenamefont{Moessner and Sondhi}(2001)}]{PhysRevB.63.224401}
\bibinfo{author}{\bibfnamefont{R.}~\bibnamefont{Moessner}} \bibnamefont{and}
  \bibinfo{author}{\bibfnamefont{S.~L.} \bibnamefont{Sondhi}},
  \bibinfo{journal}{Phys. Rev. B} \textbf{\bibinfo{volume}{63}},
  \bibinfo{pages}{224401} (\bibinfo{year}{2001}).

\bibitem[{\citenamefont{Ito et~al.}(1997)\citenamefont{Ito, Oyama, Fukaya,
  Kato, and Miura}}]{doi:10.1143/JPSJ.66.3636}
\bibinfo{author}{\bibfnamefont{A.}~\bibnamefont{Ito}},
  \bibinfo{author}{\bibfnamefont{C.}~\bibnamefont{Oyama}},
  \bibinfo{author}{\bibfnamefont{A.}~\bibnamefont{Fukaya}},
  \bibinfo{author}{\bibfnamefont{H.}~\bibnamefont{Kato}}, \bibnamefont{and}
  \bibinfo{author}{\bibfnamefont{S.}~\bibnamefont{Miura}},
  \bibinfo{journal}{Journal of the Physical Society of Japan}
  \textbf{\bibinfo{volume}{66}}, \bibinfo{pages}{3636} (\bibinfo{year}{1997}).

\bibitem[{\citenamefont{Gunnarsson et~al.}(1991)\citenamefont{Gunnarsson,
  Svedlindh, Nordblad, Lundgren, Aruga, and Ito}}]{PhysRevB.43.8199}
\bibinfo{author}{\bibfnamefont{K.}~\bibnamefont{Gunnarsson}},
  \bibinfo{author}{\bibfnamefont{P.}~\bibnamefont{Svedlindh}},
  \bibinfo{author}{\bibfnamefont{P.}~\bibnamefont{Nordblad}},
  \bibinfo{author}{\bibfnamefont{L.}~\bibnamefont{Lundgren}},
  \bibinfo{author}{\bibfnamefont{H.}~\bibnamefont{Aruga}}, \bibnamefont{and}
  \bibinfo{author}{\bibfnamefont{A.}~\bibnamefont{Ito}},
  \bibinfo{journal}{Phys. Rev. B} \textbf{\bibinfo{volume}{43}},
  \bibinfo{pages}{8199} (\bibinfo{year}{1991}),
  \urlprefix\url{https://link.aps.org/doi/10.1103/PhysRevB.43.8199}.

\end{thebibliography}

\end{document}